\documentclass[preprint]{sig-alternate-05-2015}
\usepackage{amsmath,amssymb}
\usepackage{changepage}
\DeclareMathOperator*{\argmin}{arg\,min}
\usepackage{multirow}

\makeatletter
\def\@copyrightspace{\relax}
\makeatother

\begin{document}
\title{Item-to-item recommendation based on Contextual Fisher Information} 
\numberofauthors{3}
\author{
\alignauthor B\'alint Dar\'oczy\\
       \affaddr{Inst.\ Computer Science and Control, Hungarian Academy of Sciences (MTA SZTAKI)}\\
%       \affaddr{1111 Kende street 13-17}\\
%       \affaddr{Budapest, Hungary}\\
       \email{daroczyb@ilab.sztaki.hu}
\alignauthor Frederick Ayala-G\'omez\\
       \affaddr{E\"otv\"os University}\\
%       \affaddr{1111 Kende street 13-1}\\
       \affaddr{Budapest, Hungary}\\
       \email{fayala@ilab.sztaki.hu}
\alignauthor Andr\'as A.~Bencz\'ur\\
       \affaddr{Inst.\ Computer Science and Control, Hungarian Academy of Sciences (MTA SZTAKI)}\\
%       \affaddr{1111 Kende street 13-17}\\
%       \affaddr{Budapest, Hungary}\\
       \email{benczur@sztaki.mta.hu}
}

\maketitle
\begin{abstract}

Web recommendation services bear great importance in e-commerce, as they aid the user in navigating through the items that are most relevant to her needs.
In a typical Web site, long history of previous activities or purchases by the user is rarely available.  Hence in most cases, recommenders propose items that are similar to the most recent ones viewed in the current user session. % actual != aktualis !!!
The corresponding task is called session based item-to-item recommendation.

For frequent items, it is easy to present item-to-item recommendations by ``people who viewed this, also viewed'' lists.
However, most of the items belong to the long tail, where previous actions are sparsely available.
Another difficulty is the so-called cold start problem, when the item has recently appeared and had no time yet to accumulate sufficient number of transactions.
In order to recommend a next item in a session in sparse or cold start situations, we also have to incorporate item similarity models.

In this paper we describe a probabilistic similarity model based on Random Fields to approximate item-to-item transition probabilities.  We give a generative model for the item interactions based on arbitrary distance measures over the items including explicit, implicit ratings and external metadata. The model may change in time to fit better recent events and recommend the next item based on the updated Fisher Information.

Our new model outperforms both simple similarity baseline methods and recent item-to-item recommenders, under several different performance metrics and publicly available data sets.
We reach significant gains in particular for recommending a new item following a rare item.

\end{abstract}

\section{Introduction}

Recommender systems \cite{ricci2011introduction} have become extremely common recently in a variety of areas including movies, music, news, books, and products in general.
They produce a list of recommended items by either collaborative or content based filtering.
Collaborative filtering methods \cite{amazon-recommender,sarwar01item} build models of the past user-item interactions, while content based filtering \cite{lops2011content} typically generates lists of similar items
based on item properties.

In order to assess the attitude towards the items viewed, recommender systems rely on the feedback provided by the user.
The feedback may be explicit, such as one to five stars ratings of movies in Netflix \cite{adhikari2012unreeling}. 
Most of the recommendation tasks are, however, implicit, as the user provides no like or dislike information.  
In these cases, recommenders have to rely on the implicit feedback such as time elapsed viewing an item or listening to a song.  
The Netflix Prize Challenge \cite{bennett2007netflix,koren2009bellkor} revolutionized our knowledge on recommender systems, however resulted in a bias towards explicit feedback in research.
In most Web services, the users are reluctant to create logins and prefer to browse anonymously.  Or we purchase certain types of goods, for example expensive electronics, so rarely that our previous purchases will be insufficient to create a meaningful user profile.
Several practitioners \cite{koenigstein2013towards} argue that most of the recommendation tasks they face are implicit feedback and without sufficient user history.
In \cite{pilaszy2015neighbor} the authors claim that 99\% of the recommendation industry tasks are implicit, and most of them are item-to-item. 
In these cases, we have to rely on the recent items viewed by the user in the actual shopping session.

%% Implicit recommendation can be challenging even for frequent items. The problem is even harder if we know little about the particular user who we want to recommend. This kind of ``cold start'' problem could be handled as a

In this paper we consider user independent item-to-item recommendation \cite{sarwar01item,amazon-recommender} especially in case of session recommendation. 
Best known example of this task is the Amazon list of books related to the last visited one \cite{amazon-recommender}. 
To provide the related list, the naive approach simply considers item pair frequencies.  However, for rare items, it is necessary to use global similarity data to avoid recommendations based on very low support.
In addition, we have to devise techniques that handle new items well. In the so-called cold start case \cite{schein2002methods}, the new items have yet insufficient number of interactions to reliably model their relation to the users.

Our key idea is to utilize the known, recent or popular items for item-to-item recommendation via multiple representations. The starting point of our method is the idea of \cite{koenigstein2013towards} to utilize the entire training data and not just the item-item conditional probabilities.
They propose a latent factor model for item-item pairwise relations by a low dimensional embedding.
Our new idea is to replace the ``black box''  embedding, and model item relations by a generative model defined through a Bayesian network reflecting the topology of the data set.  
As the main advantage of our method, we may use natural similarity measures compared to low dimensional embeddings, which may only use $\ell_2$ or some other vector space metric. 
Our item-to-item recommender may be based on a combination of collaborative Jaccard, cosine or any other distance, or even content, multimedia and metadata similarity.

We consider the top-$n$ recommendation task \cite{deshpande2004item}, where for each item, we have to provide a list of best next items for the session.
In this paper, just as in \cite{koenigstein2013towards}, we conducted experiments by adding 200 sampled items to the testing item to evaluate recommendations. 
However, unlike in \cite{koenigstein2013towards}, where special tricks are needed to provide nearest neighbor search due to the bias terms in their formulas, we develop a full metric space where nearest neighbor data structures can be freely used.

We evaluate our models both by the ``traditional'' top-$n$ recommendation metrics (Recall, DCG) and the recently proposed Mean Percentile Rank (MPR) \cite{hu2008collaborative}.
We observe that the two classes of metrics behave rather differently, yet our solution outperforms all baseline methods under both circumstances.

We release the source code of our method and the data preprocessing steps, including all filtering steps starting from the official data sets at 
\url{https://github.com/frederickayala/item2item_fisher}.

%https://www.dropbox.com/sh/h84pq66l3sfut5m/AAAR9yiSZnYMG7jbJZnkTq9ya?dl=0}}.

\section{Related work}

Recommender systems are surveyed in \cite{ricci2011introduction}.
A large part of recent recommender systems publications consider the Netflix Prize Competition  \cite{bennett2007netflix}, where ratings were explicitly given (1--5 stars) and the task was to predict unseen ratings.
In this paper, we consider cases where users give no explicit ratings and we have to infer their preferences from their implicit feedback \cite{koenigstein2013towards}.
Even more, we do not want to rely on rich user history of a large number of past interactions, as most recommendation tasks in the industry are called item-to-item, since the only information available is the present user session \cite{pilaszy2015neighbor}.

The first item-to-item recommender methods \cite{sarwar01item,amazon-recommender} were using similarity information to directly find nearest neighbor \cite{desrosiers2011comprehensive} transactions. 
Another solution is to extract association rules \cite{davidson2010youtube}.
Both classes of these methods deteriorate if the last item of the session is rare.

Nearest neighbor methods were criticized for two reasons.  First, the similarity metrics typically have no mathematical justification. Second, the confidence of the similarity values is often not involved when finding the nearest neighbor, which leads to overfitting in sparse cases. 
In \cite{koren2010factor}, a method is given that learns similarity weights for users, however the method gives global and not session based user recommendation.

A new method to give session recommendations was described in \cite{rendle2010factorizing} by Rendle et al.\ that models the users by factorizing personal Markov chains.
Their method is orthogonal to ours in that they provide more accurate user based models if more data is available, while we concentrate on extracting actionable knowledge from the entire data for the sparse transactions in a session.

Closest to our work is Koenigstein and Koren \cite{koenigstein2013towards}.
We use, to the greatest extent reproducible, their experimentational settings.
They resolve the sparsity problem by computing latent item factors by using all training data and representing all items in a low dimensional space.
The advantage of our item model is that we are not restricted to vector space metrics when defining the model (the low dimensional embedding, in their case).
Starting out from an arbitrary similarity definition, we may extend similarity for all items, by using all training data, in a mathematically justified way.

Finally we mention that item-to-item recommendation was also considered as a special context aware recommendation problem.
In \cite{hidasi2012fast} sequentiality as context is handled by using pairwise associations as features in an alternating least squares model by Hidasi et al.
They mention that they face the sparsity problem in setting minimum support, confidence and lift of the associations and they use category of last purchased item as fallback.
In a follow-up result \cite{hidasi2013context}, they use the same context-aware ALS algorithm, however they only consider seasonality as context in that paper.
Our result can be used independently of the ALS based methods and can easily be combined with user personalization.

\section{Item Similarity by Fisher information}

Our main result is a generative model that captures item-to-item interactions by modeling arbitrary item-item distance metrics.
We start out with a certain measure of similarity between pairs of items based possibly on implicit or explicit user feedback as well as additional, user independent metadata such as text description, linkage or even multimedia content. 
By the pairwise similarity values and potentially other model parameters $\theta$, we model item $i$ as a random variable $p (i|\theta)$. 
From $p (i|\theta)$, we will infer the distance and the conditional probability of pairs of items $i$ and $j$.

\begin{figure}
\centerline{
  \includegraphics[scale=.2]{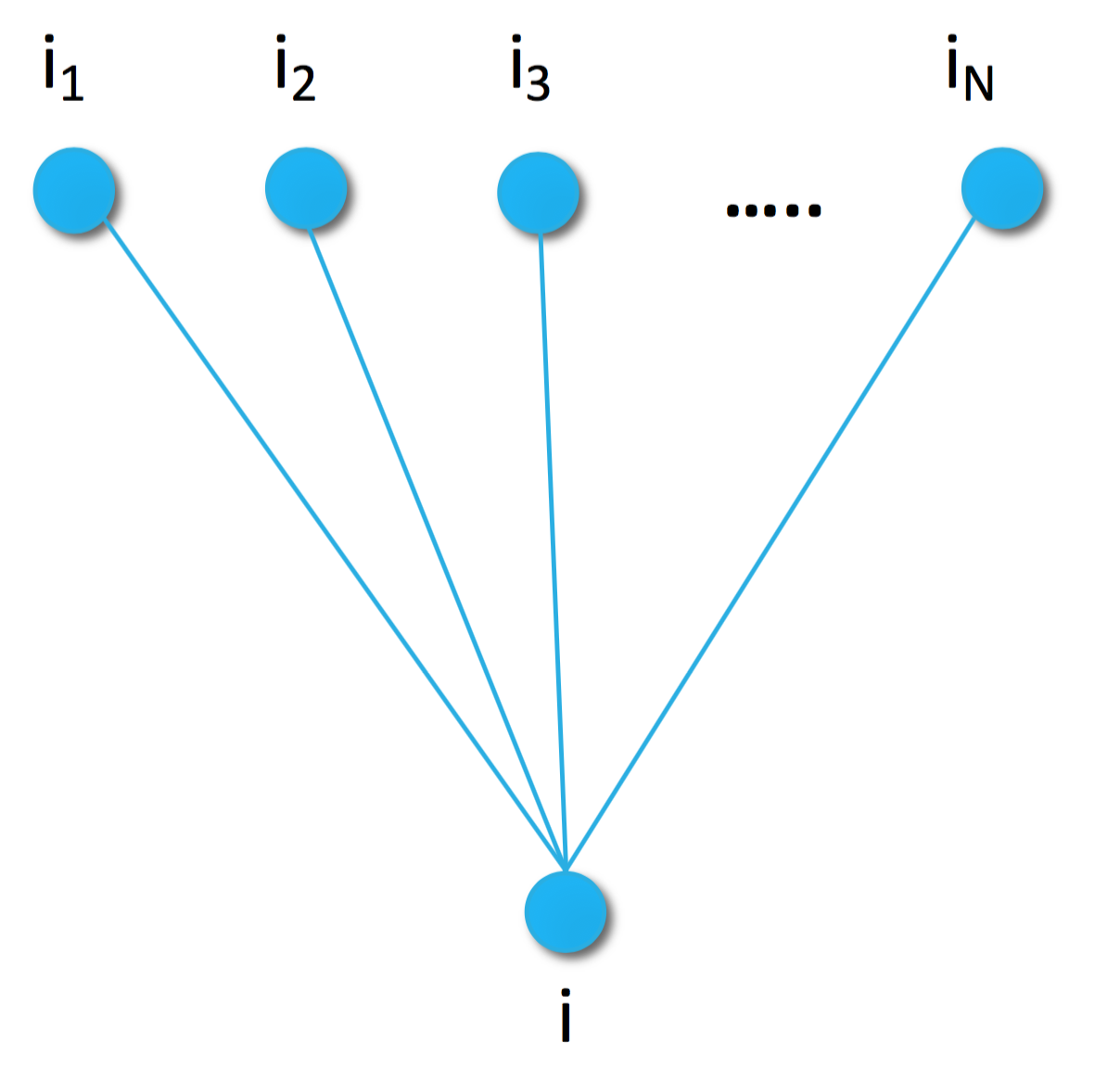}}

\caption[]{Similarity graph of item $i$ with sample items $S=\{i_1,i_2,...,i_{N}\}$ of distances $\mbox{dist}(i,i_n)$ from $i$.}
\label{fig:pairwise}
\end{figure}

Formally, let us consider a certain sample of items $S=\{i_1,i_2,\ldots,i_N\}$, and assume that we can compute the distance of any item $i$ from each of $i_n \in S$.
We will consider our current item $i$ along with its distance from each $i_n\in S$ as a random variable generated by a Markov Random Field.
For example, the simplest Markov Random Field can be obtained by using a graph with edges between item $i$ and items $i_n\in S$, as shown in  Fig.~\ref{fig:pairwise}. 

Let us assume that we are given a Markov Random Field generative model for $p (i|\theta)$.
By the Hammersley-Clifford theorem \cite{hammersley1971markov}, the joint distribution of $p (i|\theta)$ is then a Gibbs distribution. 
As the theorem states in more detail, given the potential function over the maximal cliques of the random field, the joint distribution of the generative model for $i$ is a Gibbs distribution of form
\begin{equation}
    p(i \mid \theta) = {\mathrm{e}^{-U(i\mid \theta)}}/{Z(\theta)}
    \label{eq:gibbs}
\end{equation}
where $U(i\mid \theta)$ is the energy function and 
\begin{equation}
Z(\theta) = \sum_{i} \mathrm{e}^{-U(i \mid \theta)}
\nonumber
\end{equation}
is the expected value of the exponent of the energy function over our generative model, a normalization term called the partition function.  If the model parameters are previously determined, then $Z(\theta)$ is a constant. 

Given a Markov Random Field defined by a certain graph such as the one in Fig.~\ref{fig:pairwise} (or some more complex graph defined later), a wide variety of proper energy functions can be used to define a Gibbs distribution. The weak but necessary restrictions are that the energy function has to be positive real valued, additive over the maximal cliques of the graph, and more probable configurations (specific sets of parameters) have to have lower energy. 

Now we define the simplest similarity graph seen in Fig.~\ref{fig:pairwise} as follows.  Let a finite sample set $S=\{i_1,..,i_{N}\}$, and a distance (or divergence) defined over any item pairs be given.  Since all the edges are between the elements of the sample set and the particular item $i$, we may formulate the energy function for \eqref{eq:gibbs} as
\begin{equation}
U(i \mid \theta=\{\alpha_1,..,\alpha_{N}\}) := \sum_{n=1}^{N} \alpha_n \mbox{dist}(i,i_n),
\label{eq:potential}
\end{equation}
where $\theta$ is the hyperparameter set defined over the elements in the sample set.

\begin{figure}
\centerline{
  \includegraphics[scale=.2]{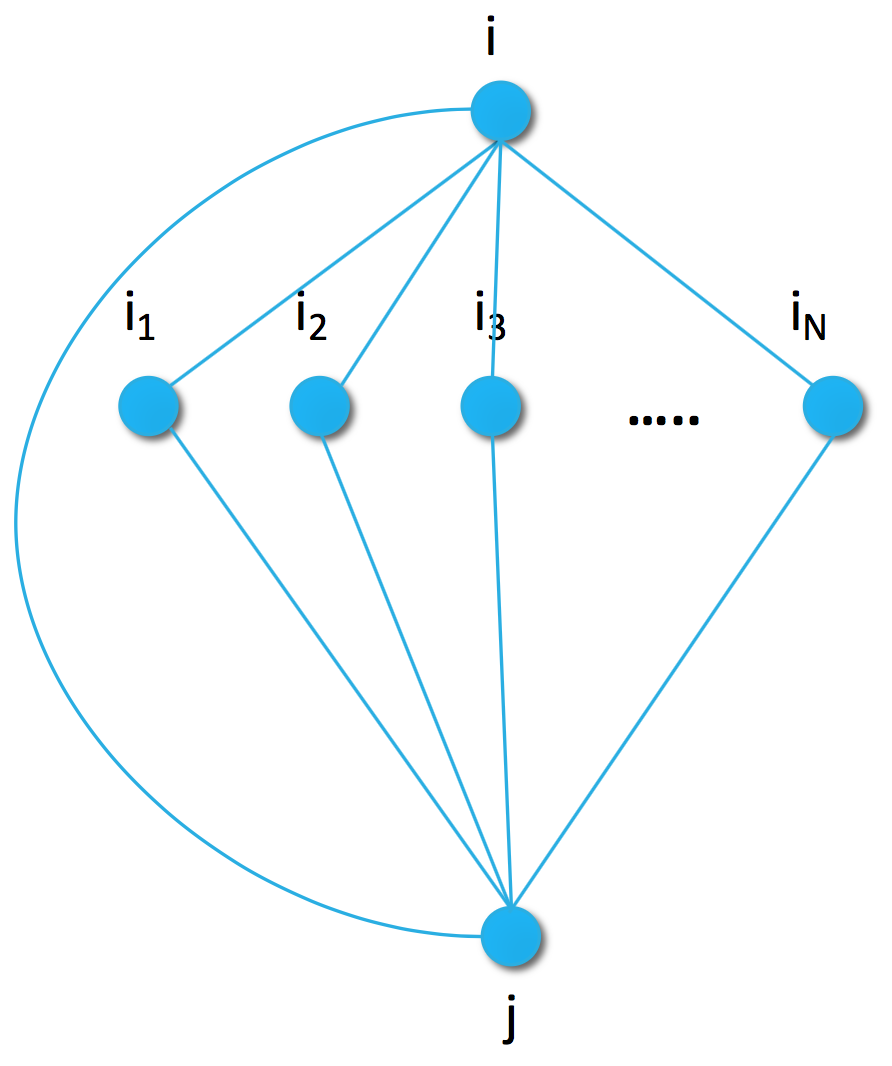}}

\caption[]{Pairwise similarity graph with sample set $S=\{i_1,i_2,...,i_{N}\}$ for a pair of items $i$ and $j$.}
\label{fig:join}
\end{figure}

In a more complex model, we capture the connection between pairs of items by extending the generative graph model with an additional node for the previous item as shown in Fig.~\ref{fig:join}.  In the pairwise similarity graph, the maximal clique size increases to three. To capture the joint energy, we can use a heuristic approximation similar to the pseudo-likelihood method \cite{besag1975statistical}: we approximate the joint distribution of each size three clique as the sum of the individual edges, as follows:
\begin{equation}
\label{eq:potential_joined}
U(i,j \mid \theta) := \sum_{n=1}^{N} \beta_{n} (\mbox{dist}(i,i_n) + \mbox{dist}(j,i_n) + \mbox{dist}(i,j)),
\end{equation}
where $\theta = \{\beta_n\}$.

At first glance, the additive approximation seems to oversimplify the clique potential and falls back to the form of equation \eqref{eq:potential}. However, the effect of the clique is apparently captured by the common clique hyperparameter $\beta_{n}$, as also confirmed by our experimental results.

\subsection{The Fisher Information}

Let us consider a general parametric class of probability models $p(i| \theta)$, where $\theta\in \Theta \subseteq \mathbb{R}^\ell$ lies in a space as defined for example by the similarity graphs of equation~\eqref{eq:gibbs}, of some positive integer dimension $\ell$. The collection of models with parameters from a general hyperparameter space $\Theta$ can then be viewed as a (statistical) manifold $M_\Theta$, provided that the dependence of the potential on $\Theta$ is sufficiently smooth. By \cite{Jo}, $M_\Theta$ can be turned into a Riemann manifold by giving an inner product (kernel) at the tangent space of each point $p(i| \theta) \in M_\Theta$, where the inner product varies smoothly with $p$. 

The notion of the inner product over $p(i| \theta)$ allows us to define the so-called Fisher metric on $M$. The fundamental result of \u{C}encov \cite{cencov1982} states that the Fisher metric exhibits a unique invariance property under some maps which are quite natural in the context of probability. Thus, one can view the use of Fisher kernel as an attempt to introduce a natural comparison of the items on the basis of the generative model \cite{JH}. 

We start defining the Fisher kernel over the manifold $M_\Theta$ of probabilities $p(i| \theta)$ as in equation~\eqref{eq:gibbs} by considering the tangent space.
The tangent vector 
\begin{equation}
G_i=\nabla_\theta \log p(i|\theta)=\left(
\frac{\partial}{\partial \theta_1} \log p(i|\theta), \ldots,
\frac{\partial}{\partial \theta_l} \log p(i|\theta)\right)
\label{eq:Fisher-score}
\end{equation}
is called the {\em Fisher score} of item $i$. 
The {\em Fisher information matrix} is a positive semidefinite matrix defined as
\begin{equation}
F(\theta):= {\bf E_{\theta}}(\nabla _\theta\log p(i|\theta)\nabla _\theta \log p(i|\theta)^T),
\label{eq:fi}
\end{equation}
%
%Now the mapping $i\mapsto \phi_i $ of items to feature vectors can be $i\mapsto F^{-\frac 12}G_i$ (we suppressed here the dependence on $\theta$), the {\em Fisher vector} where $F(\theta)$ is a positive semidefinite matrix, which varies smoothly with the base point $\theta$. Such positive semidefinite matrices are provided by the Fisher information matrix
%
%
where the expectation is taken over $p(i|\theta)$. In particular, if $p(i|\theta)$ is a probability density function, then the $nm$-th entry of 
$F(\theta)$  is 
$$ F_{nm}=\sum_{i} p(i|\theta)
\left(\frac{\partial}{\partial \theta_n} \log p(i|\theta)\right)
\left(\frac{\partial}{\partial \theta_m} \log p(i|\theta)\right).$$

Thus, to capture the generative process, the gradient space of $M_\Theta$ is used to derive the Fisher vector, a mathematically grounded feature representation of item $i$. The corresponding kernel function 
\begin{equation}
K(i,j):= G^T_iF^{-1}G_j
\label{eq:Fisher-kernel}
\end{equation}
is called the {\em Fisher kernel}. An intuitive interpretation is that $G_i$ gives the direction where the parameter vector $\theta$ should be changed to fit item $i$ the best \cite{perronnin2007fisher}.

\subsection{Item-Item Fisher Conditional Score (FC)}
\label{sec:fs_rank}
\label{sect:FC}

Our first item-to-item recommender method will involve similarity information in the item-item transition conditional probability computation by using Fisher scores as in equation~\eqref{eq:Fisher-score}.
By the Bayes theorem,
\begin{equation*}
G_{j|i} = \nabla_{\theta} \log p(j \mid i; \theta) = \nabla_{\theta} \log \frac{p(i,j \mid \theta)}{p(i \mid \theta)}
\nonumber
\end{equation*}
\begin{equation}
= \nabla_{\theta} \log p(i,j \mid \theta) - \nabla_{\theta} \log p(i \mid \theta),
\label{eq:fisher_score_cond}
\end{equation}
thus we need to determine the joint and the marginal distributions for a particular item pair. 

First, let us calculate the Fisher score of \eqref{eq:Fisher-score} with $p (i|\theta)$ of the single item generative model defined by \eqref{eq:potential}, 
\begin{equation}
\begin{split}
    G_i^k(\theta) &=\nabla_{\theta_k} \log p(i|\theta)\\
          &= \frac{1}{Z(\theta)}\sum_{i} \mathrm{e}^{-U(i \mid \theta)} \frac{\partial{U(i \mid \theta)}}{\partial{\theta_k}} - \frac{\partial{U(i \mid \theta)}}{\partial{\theta_k}} \\
          &= \sum_{i} \frac{\mathrm{e}^{-U(i \mid \theta)}}{Z(\theta)} \frac{\partial{U(i \mid \theta)}}{\partial{\theta_k}} - \frac{\partial{U(i \mid \theta)}}{\partial{\theta_k}}.
\end{split}
\nonumber
\end{equation}
By \eqref{eq:gibbs}, our formula can be simplified as 
\begin{equation}
\begin{split}
    G_i^k(\theta) &= \sum_{i} p(i \mid \theta) \frac{\partial{U(i \mid \theta)}}{\partial{\theta_k}} - \frac{\partial{U(i \mid \theta)}}{\partial{\theta_k}} \\
    				 &= {\textstyle {\bf E_{\theta}}[\frac{\partial{(U(i \mid \theta)}}{\partial{\theta_k}}] - \frac{\partial{U(i \mid \theta)}}{\partial{\theta_k}}.}
\end{split}
\label{eq:gix}
\end{equation}
For an energy function as in equation \eqref{eq:potential}, the Fisher score of $i$ has a simple form,
\begin{equation}
    G_i^k(\theta)= {\bf E_{\theta}}[\mbox{dist}(i,i_k)] - \mbox{dist}(i,i_k),
\label{eq:fisher_pair}
\end{equation} 
and similarly for equation \eqref{eq:potential_joined}, 
\begin{equation}
\begin{split}
    G_{ij}^{k}(\theta) = {\bf E_{\theta}}[\mbox{dist}(i,i_k) + \mbox{dist}(j,i_k)+\mbox{dist}(i,j)] \\
    - (\mbox{dist}(i,i_k)+ \mbox{dist}(j,i_k)+\mbox{dist}(i,j)). 
\end{split}
\label{eq:fisher_joined}
\end{equation} 

Now, if we put \eqref{eq:fisher_pair} and \eqref{eq:fisher_joined} into \eqref{eq:fisher_score_cond}, several terms cancel out and the Fisher score has the simple form
\begin{equation}
G_{j|i}^k = {\bf E_{\theta}}[\mbox{dist}(j,i_k) + \mbox{dist}(i,j)] - (\mbox{dist}(j,i_k) + \mbox{dist}(i,j)).
\nonumber
\end{equation}
The above formula involves the distance values on the right side, which are readily available, and the expected values on the left side, which may be estimated by using the training data. We note that here we make a heuristic approximation: instead of computing the expected values e.g.\ by simulation, we substitute the mean of the distances from the training data.

As we discussed previously, the Fisher score resembles how well the model can fit the data, thus we can recommend the best fitting next item $j^{*}$ based on the norm of the Fisher score,
\begin{equation}
j^{*} = \argmin_{j \neq i} || G_{j|i}(\theta) ||,
\nonumber
\end{equation}
where we will use $\ell_2$ for norm in our experiments.

\subsection{Item-Item Fisher Distance (FD)}
\label{sect:FD}

In our second model, we rank the next item by its distance from the last one, based on the Fisher metric. With the Fisher kernel $K(i,j)$, the \emph{Fisher distance} can be formulated as
\begin{equation}
\mbox{dist}_F(i,j) = \sqrt{K(i,i) - 2 K(i,j) + K(j,j)},
\label{eq:Fisher-dist}
\end{equation}   
thus we need to compute the Fisher kernel over our generative model as in \eqref{eq:Fisher-kernel}.  The computational complexity of the Fisher information matrix estimated on the training set is $\mathcal{O}(T |\theta|^2)$, where $T$ is the size of the training set.  To reduce the complexity to $\mathcal{O}(T |\theta|)$, we can approximate the Fisher information matrix with the diagonal as suggested in \cite{JH,perronnin2007fisher}.
Hence we will only use the diagonal of the Fisher information matrix, 
\begin{equation}
\begin{split}
    F_{k,k} &={\bf E_{\theta}}[\nabla_{\theta_k} \log p(i|\theta)^T \nabla_{\theta_k} \log p(i|\theta)] \\
    &={\bf E_{\theta}}[({\bf E_{\theta}}[\frac{\partial{U(i \mid \theta)}}{\partial{\theta_k}}] - \frac{\partial{(U(i \mid \theta)}}{\partial{\theta_k}})^2] . \\
\end{split}
\nonumber
\end{equation}

For the energy functions of equations \eqref{eq:potential} and \eqref{eq:potential_joined}, the diagonal of the Fisher kernel is the standard deviation of the distances from the samples. We give the Fisher vector of $i$ for 
\eqref{eq:potential}:
\begin{equation}
\begin{split}    
    \mathcal{G}_i^k &= F^{-\frac 12} G_i^k \approx F_{kk}^{-\frac 12} G_i^k\\
    &=\frac {{\bf E_{\theta}}[\mbox{dist}(i,i_k)] - \mbox{dist}(i,i_k)} {{\bf E_{\theta}^{\frac 12}}[({\bf E_{\theta}}[\mbox{dist}(i,i_k)] - \mbox{dist}(i,i_k))^2]}.
    \nonumber
\end{split}
\label{eq:kernel}
\end{equation}

The final kernel function is
\begin{equation}
\begin{split}
K(i,j) &= G_i^T F^{-1} G_j \approx G_i^T F_{diag}^{-1} G_j \\
	   &= G_i^T F_{diag}^{-\frac 12} F_{diag}^{-\frac 12} G_j = \sum_k \mathcal{G}_i^k \mathcal{G}_j^k.
\end{split}
\nonumber
\end{equation}
By substituting into \eqref{eq:Fisher-dist}, the recommended next item after item $i$ will be
\begin{equation}
j^{*} = \argmin_{j \neq i} \text{dist}_F(i,j).
\nonumber 
\end{equation}

\subsection{Multimodal Fisher probabilities and distances}
\label{sec:fish_comb}

\begin{figure}
\centerline{
\includegraphics[scale=.2]{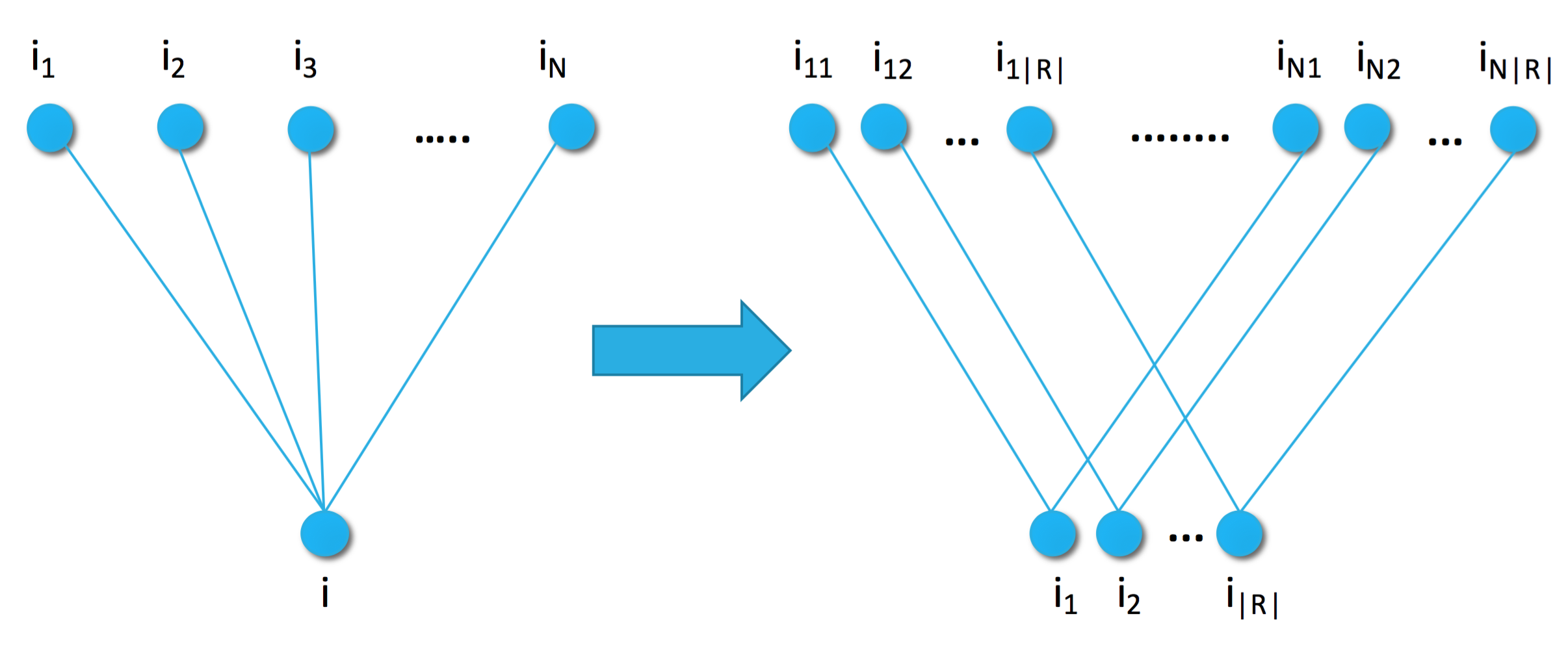}}
\caption{The single and multimodal similarity graph with sample set $S=\{i_1,i_2,...,i_{N}\}$ and $|R|$ modalities.}
\label{fig:multi}
\end{figure}

So far we considered only a single distance or divergence measure over the items. We may expand the model with additional distances with a simple modification to the graph of Fig.~\ref{fig:pairwise}. 
We expand the points of the original graph into new points $R_i=\{r_{i,1},..,r_{i,|R|}\}$ corresponding to $R$ representatives for each item $i_n$ in Fig.~\ref{fig:multi}. There will be an edge between two item representations $r_{i,\ell}$ and $r_{j,k}$ if they are the same type of representation ($\ell=k$) and the two item was connected in the original graph. This transformation does not affect the the maximal clique size and therefore the energy function is a simple addition, as
\begin{equation}
U(i \mid \theta) = \sum_{n=1}^{N} \sum_{r=1}^{|R|} \alpha_{nr} \mbox{dist}_r(i_r,i_{nr}),
\label{eq:potential_multi}
\end{equation}
and if we expand the joint similarity graph to a multimodal graph, the energy function will be
\begin{equation}
\begin{split}
\label{eq:potential_joined_multi}
U(i,j \mid \theta) = \sum_{n=1}^{N} \sum_{r=1}^{|R|} \beta_{nr} (\mbox{dist}_r(i_r,i_{nr}) \\
+ \mbox{dist}_r(j_r,i_{nr}) + \mbox{dist}_r(i_r,j_r)).
\end{split}
\end{equation}

Now, let the Fisher score for any distance measure $r \in R$ be $G_{ir}$, than the Fisher score for the multimodal graph is concatenation of the unimodal Fisher scores as
\begin{equation}
G_i^{multi} = \{G_{i1},..,G_{i|R|}\},
\nonumber
\end{equation}
and therefore the norm of the multimodal Fisher score is a simple sum over the norms: 
\begin{equation}
|| G_i^{multi}|| = \sum_{r=1}^{|R|} ||G_{ir}||.
\label{eq:fs_multi}
\end{equation}
The calculation is similar for the Fisher kernel of equation \eqref{eq:kernel}, thus the multimodal kernel can be expressed as
\begin{equation}
K_{multi}(i,j) = \sum_{r=1}^{|R|} K_r(i,j).
\label{eq:fk_multi}
\end{equation}

\section{Similarity Measures}

Next we enumerate distance and divergence measures that can be used in the energy functions \eqref{eq:potential} and \eqref{eq:potential_joined}. 
Without using the Fisher information machinery, these measures yield the natural baseline methods for item-to-item recommendation.
We list both implicit feedback collaborative filtering and content based measures.

\subsection{Feedback Similarity}

For user implicit feedback on item pairs, various joint and conditional distribution measures can be defined based on the frequency $f_i$ and $f_{ij}$ of items $i$ and item pairs $i,j$, as follows.
\begin{enumerate}
\item Cosine similarity (Cos):
\begin{equation}
cos(i,j) = \frac{f_{ij}}{f_i f_j}.
\nonumber
\end{equation}
\item Jaccard similarity (JC):
\begin{equation}
JC(i,j) = \frac{f_{ij}}{f_i+f_j-f_{ij}}.
\nonumber
\end{equation}
\item Empirical Conditional Probability (ECP): estimates the item transition probability:
\begin{equation}
ECP(j|i) = \frac{f_{ij}}{f_i+1},
\nonumber 
\end{equation}
where the value 1 is a smoothing constant. 
\end{enumerate}
Additionally, in \cite{koenigstein2013towards} the authors suggested a model, the Euclidean Item Recommender (EIR) to approximate the transition probabilities with the following conditional probability
\begin{equation}
p(j | i) = \frac{\exp^{-||x_i-x_j||^2 + b_j}}{\sum_{k \neq i} \exp^{-||x_i-x_k||^2+b_k}},
\nonumber
\end{equation}
where they learn a latent vector $x_i$ and bias $b_i$ for item $i$. 

All of the above measures can be used in the energy function as the distance measure after small modifications. 

Now, let us assume that our similarity graph (Fig.~\ref{fig:pairwise}) has only one sample element $i$ and the conditional item is also $i$. The Fisher kernel will be,
\begin{equation}
\nonumber
\begin{split}
K(i,j) &= \frac{1}{\sigma_{i}^2} (\mu_i - \mbox{dist}(i,i))(\mu_i-\mbox{dist}(i,j)) \\
       &= \frac{\mu_i^2}{\sigma_{i}^2} - \frac{\mu_i}{\sigma_{i}^2} \mbox{dist}(i,j)) \\
       &= C_1-C_2*\mbox{dist}(i,j),
\end{split}
\end{equation}

where $\mu_i$ and $\sigma_i$ are the expected value and variance of distance from item $i$. Therefore if we fix $\theta$, $C_1$ and $C_2$ are positive constants and the minimum of the 
Fisher distance will be %???

\begin{equation}
\nonumber
\begin{split}
\min_{j \neq i} \text{dist}_F(i,j) &= \min_{j \neq i} \sqrt{K(i,i) - 2 K(i,j) + K(j,j)} \\
                   &= \min_{j \neq i} \sqrt{2C_2*\mbox{dist}(i,j)} = \min_{j \neq i} \mbox{dist}(i,j).
\end{split}
\end{equation}

Hence if we measure the distance over the latent factors of EIR, the recommended items will be the same as defined by EIR (equation~10 in \cite{koenigstein2013towards}).

\subsection{Content Similarity}

We may use item metadata for measuring similarity based on content.
For example, we may use the semantic structure of the items based on DBPedia\footnote{\url{http://wiki.dbpedia.org}} \cite{auer2007dbpedia} by collecting the 
Resource Description Framework (RDF) nodes and links or other semantic knowledge graphs corresponding to the items to be recommended.  % ???
In case of movies, for example, the extracted graph is an entity tree including directors, actors, genre, topics, etc. We have multiple options to define  similarity based on the semantic graph.
We may use the Jaccard similarity or the Jensen-Shannon divergence of the bag of nodes in the semantic graph, or even use Graph kernels \cite{losch2012graph}.

%Let $P$ and $Q$ be the probability distributions for item $i$ and $j$, then the Jensen-Shannon divergence is
%
%\begin{equation}
%\nonumber
%JS(P,Q) = \frac{1}{2} (KL(P||M)+KL(Q||M))
%\end{equation}
%
%where $M=\frac{1}{2}(P+Q)$ and $KL(P||Q)=\sum_i P_i \log \frac{P_i}{Q_i}$ is the Kullback-Leibler divergence. 

\section{Experiments}

We performed experiments on four publicly available data sets.
As baseline methods, we computed four item-item similarity measures: Empirical Conditional Probability (ECP), Cosine 
(Cos), Jaccard (JC) as defined in the Traditional Similarity Metric Section, 
and we also implemented the Euclidean Item Recommender of \cite{koenigstein2013towards}. 
As content similarity we computed Jaccard similarity. % and Jensen-Shannon divergence. % ??? for content-based recommendation ?
For evaluation, we use Mean Percentile Rank (MPR) \cite{koenigstein2013towards}, Recall, and Discounted Cumulative Gain (DCG), which we define next.

\begin{figure}
\centerline{
  \includegraphics[scale=.2]{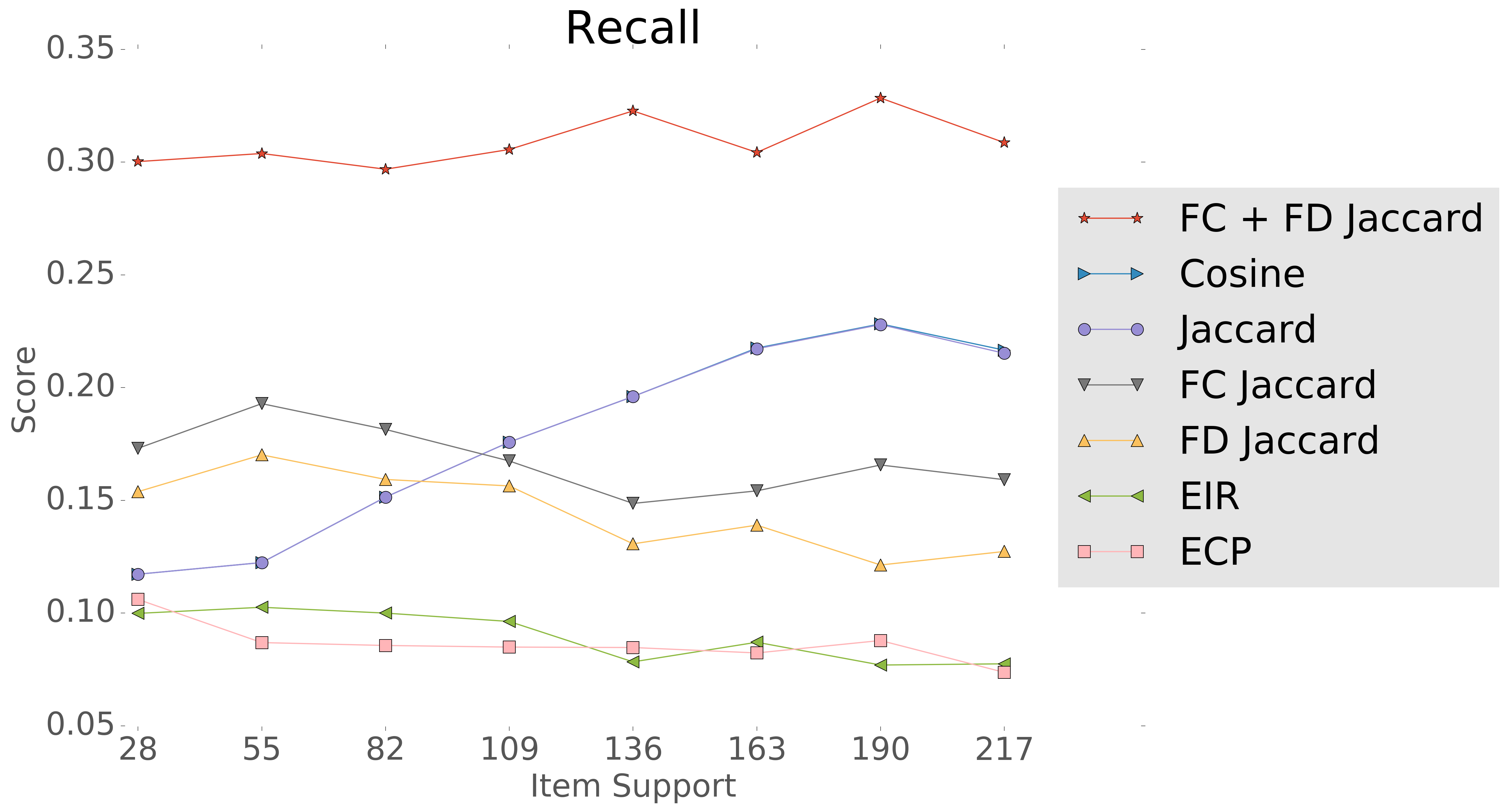}}

\caption[]{Recall@20 as the function of item support for the Netflix data set.
FD + FC stands for the linear combination of the scores of FD and FD Jaccard.}
\label{fig:supp_netf}
\end{figure}

\begin{figure*}[t]
\centerline{
\includegraphics[scale=.32]{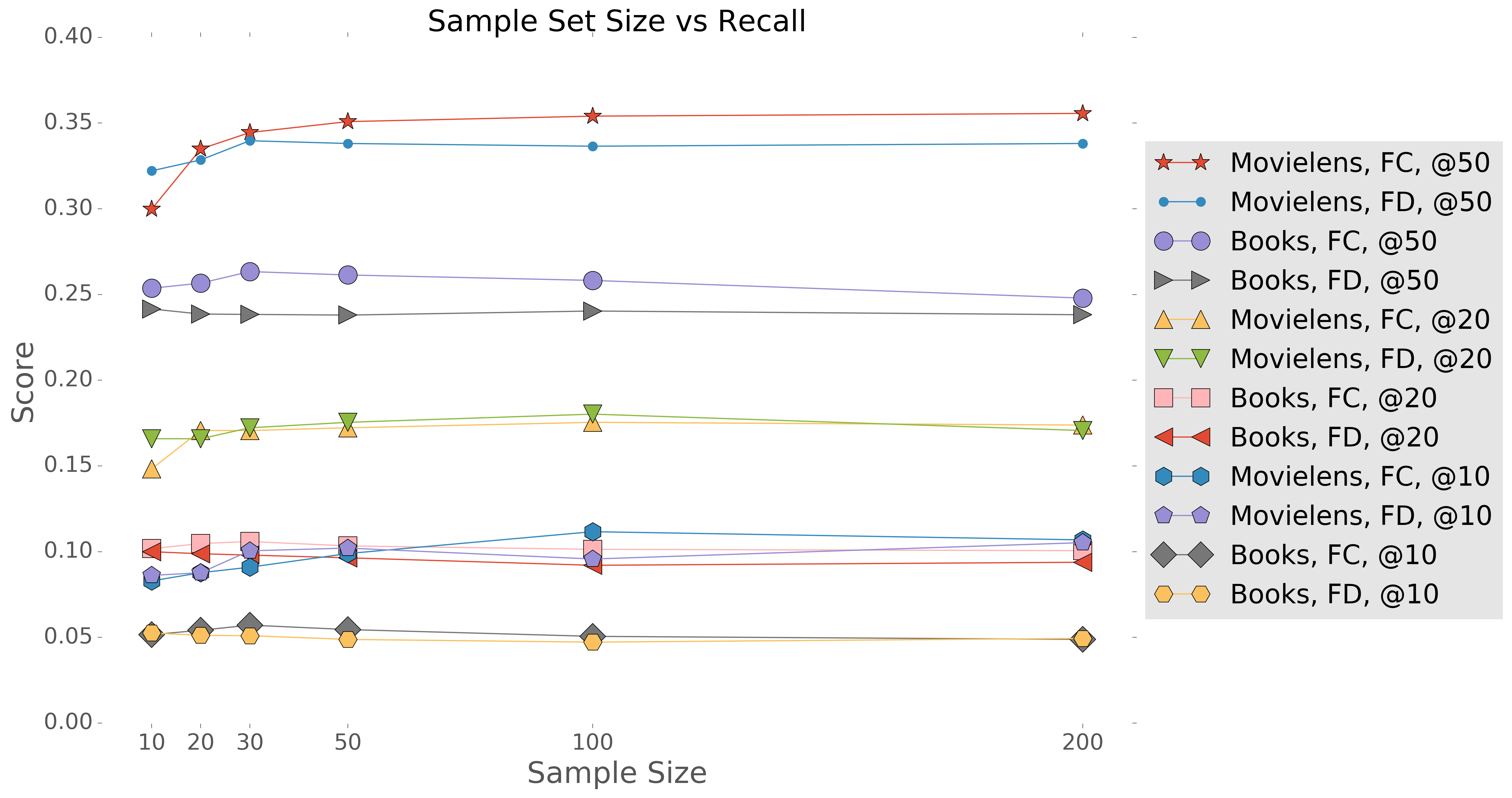}}
\centerline{
\includegraphics[scale=.32]{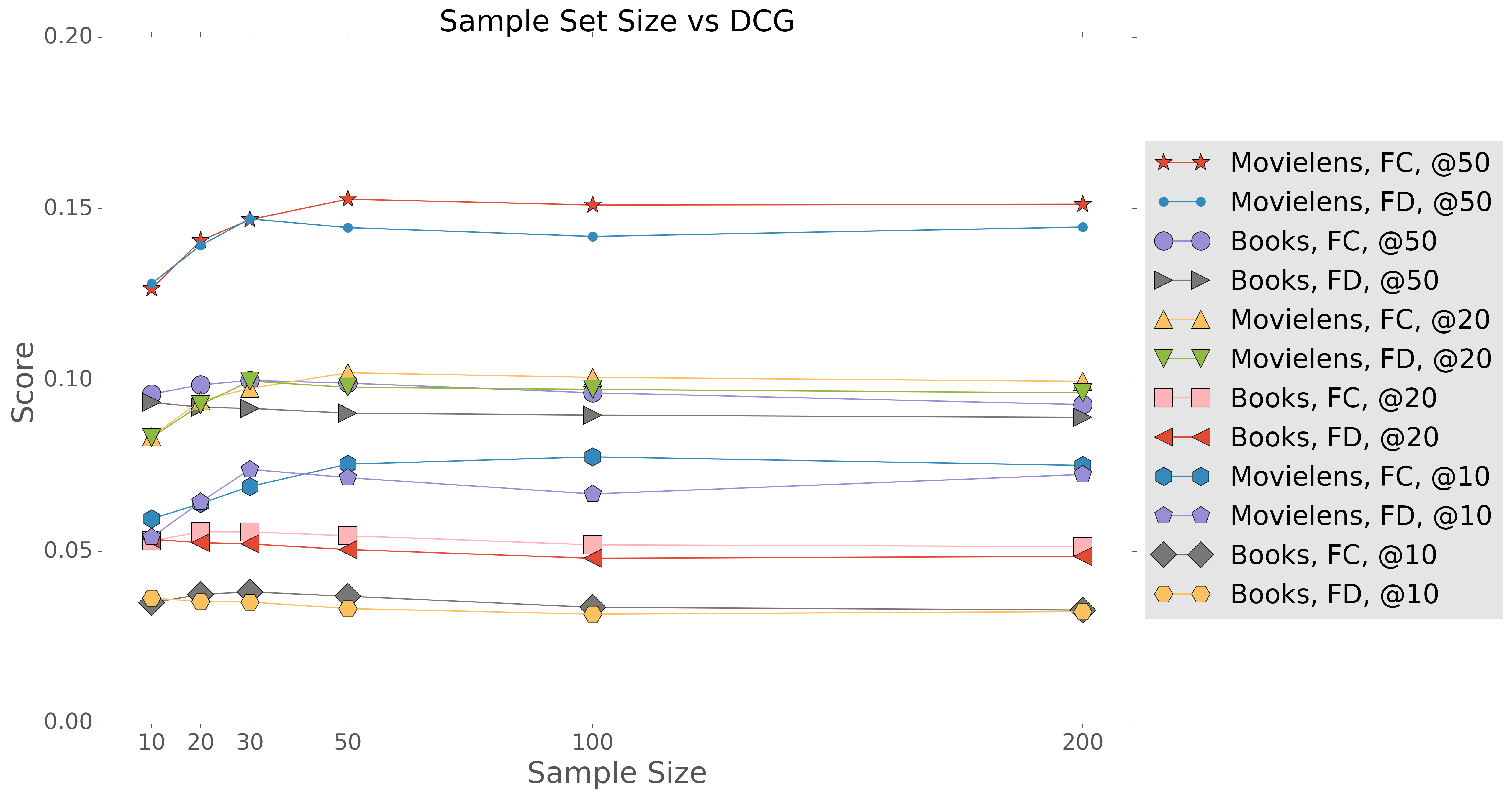}}
\centerline{
  \includegraphics[scale=.32]{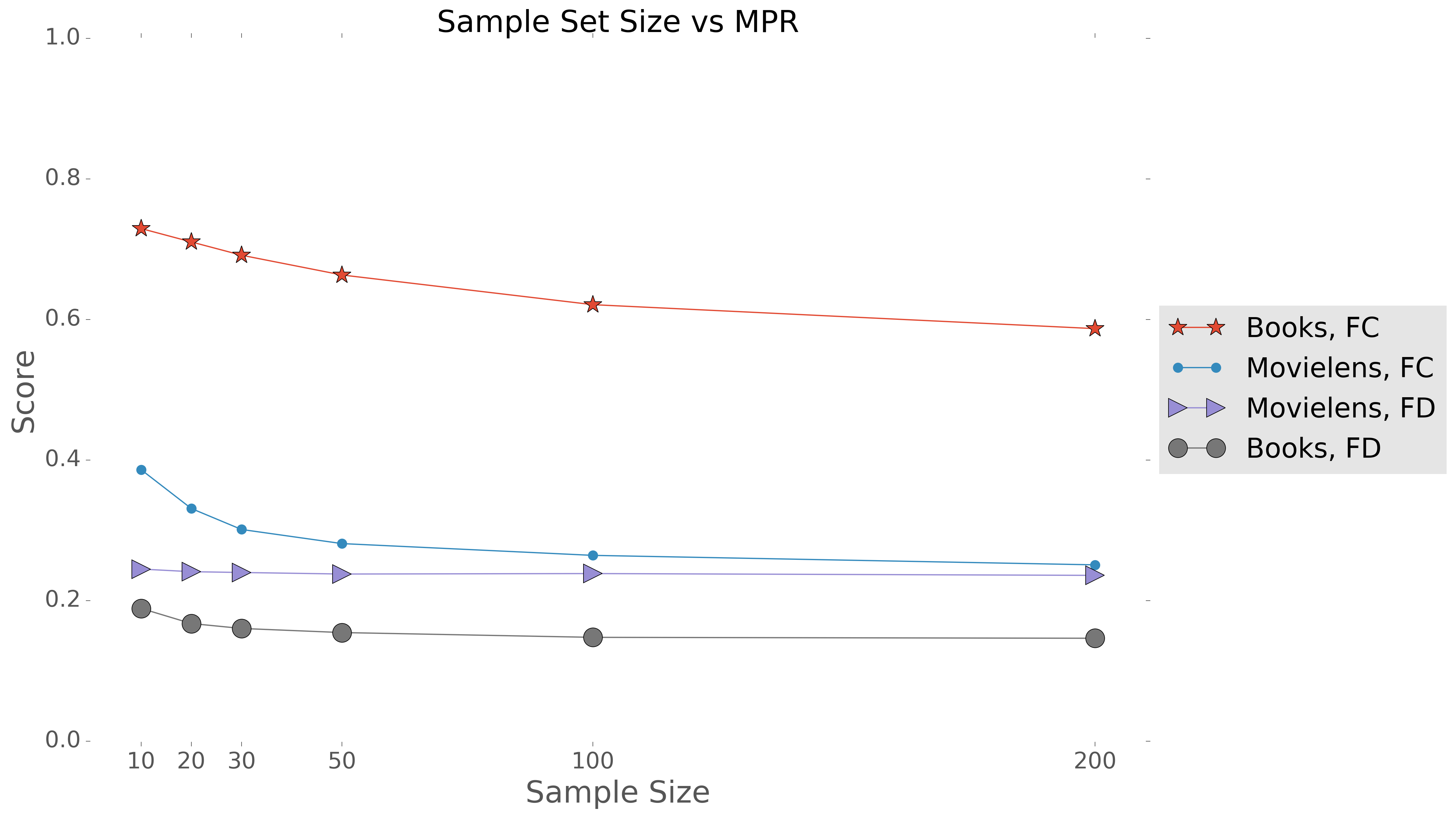}}

\caption[]{Recall, DCG and MPR as the function of the sample set size.
  FC and FD are the Fisher information methods of Sections~\ref{sect:FC} and~\ref{sect:FD}.
  Both methods are defined over the item implicit feedback Jaccard similarity.}
\label{fig:samples}
\end{figure*}

\subsection{Evaluation Metrics}

Let us consider an item $i$ and a ranking $r_1,r_2,\ldots$ of the possible next items.
We define Recall@K as a binary value: if the actual next item $j$ is ranked K or higher, Recall@K is one, otherwise zero:
\begin{equation}
\mbox{Recall@}K(r_1,\ldots,r_K) = \sum_{\ell=1}^K \delta_{r_\ell,j},
\nonumber
\end{equation}
where $\delta_{r_\ell,j}$ is the Kronecker delta: one if $r_\ell=j$ and zero otherwise. Similarly, DCG@K is defined to be zero if the test 
item j is not in the top K recommended items, otherwise
\begin{equation}
\mbox{DCG@}K(r_1,\ldots,r_K) = \sum_{\ell=1}^K \frac{\delta_{r_\ell,j}}{\log_2(\ell+1)}.
\nonumber
\end{equation} 

%Worth to mention, the presentation of the recommended items may effect the performance.

In \cite{koenigstein2013towards}, the authors measure item-to-item recommendation by mean percentile rank (MPR) as defined next.
Given the ranked list $r_1,r_2,\ldots$ and the actual next item $j$ of rank $k$, $j=r_k$, percentile rank is defined as %???  or -> of cahnged by Adri
\begin{equation}
\text{PR}(j) = \sum_{\ell > k} f_\ell / \sum_\ell f_\ell.
\nonumber
\end{equation}
The authors propose a slight modification if certain elements are ranked in tie with $j$,
\begin{equation}
\text{PR}(j) = \frac{\sum_{\text{$\ell$ ranked strictly after $j$}} f_\ell + 0.5*\sum_{\text{$\ell$ ranked in tie with $j$}} f_\ell} {\sum_\ell f_\ell}.
\nonumber
\end{equation}
Finally, MPR is the mean PR over all testing events.
Unlike Recall and DCG where higher value indicates better performance, for MPR, the lower the better.

The main difference of MPR compared to DCG and Recall is that MPR also takes the  popularity of the items into consideration.
Since the occurrence of the items is typically non-uniform, if a popular item surpasses the actual next item, the penalty in PR could be high.
On the other hand, DCG and Recall give more emphasis to the accuracy of the top recommended items.
As we will see in the experiments, the relation of Recall or DCG with MPR depends on the distribution of the conditional items. 

\begin{table}[t]
\caption[]{Data sets used in the experiments.}
\centerline{
\begin{tabular}{|l|r|r|r|r|r|}\hline
Data set       & Items  & Users  & Training & Testing\\
               &        &	 & pairs    & pairs \\ \hline
Netflix        & 17749  & 478488 & 7082109  & 127756 \\
MovieLens      & 3683   & 6040   & 670220   & 15425 \\
Yahoo!\ Music  & 433903 & 497881 & 27629731 & 351344\\
Books      & 340536 & 103723 & 1017118  & 37403 \\\hline
\end{tabular}}
\label{tab:datasets}
\end{table}

\subsection{Data sets and Experimental Settings}

We carried out experiments over four data sets: Netflix \cite{bennett2007netflix}, MovieLens\footnote{\url{http://grouplens.org/datasets/movielens/}}, 
Books \cite{ziegler2005improving} and Yahoo!\ Music \cite{dror2012yahoo}.

Our evaluation methodology was similar to \cite{koenigstein2013towards}.
For each user, we made a random ordering over the items and split the list into two subsets.
The number of training and testing pairs and the properties of the data sets can be seen in Table~\ref{tab:datasets}.
During testing the evaluation was performed over a sampled set of 200 items as in \cite{koenigstein2013towards} for all three metrics and solved ties arbitrary in the ranking except 
in case of MPR. % because the expected value of the cummulative frequency of surprassing items will be the same.

Note that in our experiments, all algorithms use the item frequencies of the training period as input parameters.
It is however possible to keep the current frequencies up to date and recalculate the prediction of each algorithm on the fly.

\begin{figure}
\centerline{
  \includegraphics[scale=.21]{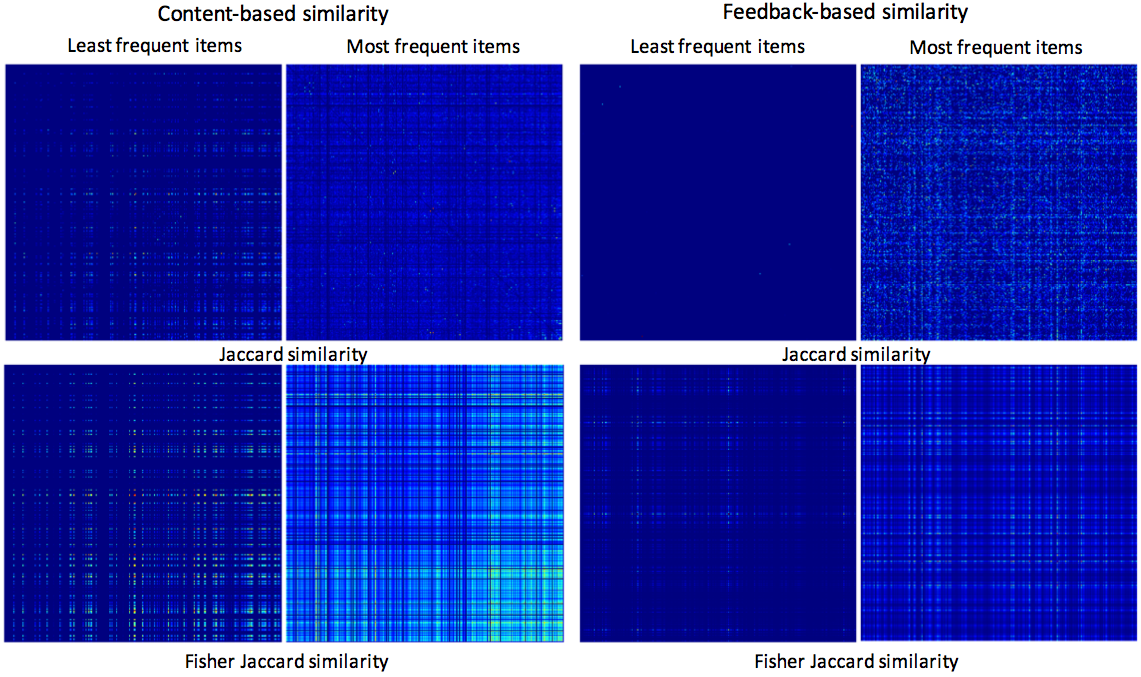}}
\caption[]{Density of different similarity measures between the least frequent items and between the most frequent items on MovieLens dataset. Note we set the diagonal to zero to highlight non diagonal entries.}
\label{fig:sparsity}
\end{figure}

\subsection{Experimental Results}

\newcommand\baslin[1]{#1}
\newcommand\bestal[1]{\textbf{#1}}
\begin{table} 
  \caption[]{Experimentation summary with the best performing collaborative baseline and new method names and performance metrics shown by filtering items at three different maximum frequency thresholds.  For most methods, there are (up to rounding errors) two best baseline and another two best new method candidates, except for Recall where a third method Cosine appears in the cell marked by a star ($*$).
%The full experimentation tables are included in the supplementary material.
}
\centering
\begin{tabular}{rlrrrrr} \hline
  &\multicolumn{1}{p{2cm}}{\scriptsize{Best baseline\& new method}}
  &\multicolumn{1}{p{0.6cm}}{\scriptsize{Max freq}}
  &\multicolumn{1}{p{0.8cm}}{\scriptsize{Movie\-Lens}}
  &\multicolumn{1}{p{0.8cm}}{\scriptsize{Books}}
  &\multicolumn{1}{p{0.8cm}}{\scriptsize{Yahoo!\ Music}}
  &\multicolumn{1}{p{0.8cm}}{\scriptsize{Net\-flix}} \\ \hline 
         \multirow{7}{*}{{\rotatebox[origin=c]{90}{MPR}}}
& EIR            & 25\% &\baslin{0.33}&\baslin{0.48} &\baslin{0.24}&\baslin{0.34} \\
&                & 50\% &\baslin{0.35}&\baslin{0.48} &\baslin{0.25}&\baslin{0.35} \\
&                & 75\% &\baslin{0.36}&\baslin{0.47} &\baslin{0.25}&\baslin{0.38} \\ \cline{2-7}
& FD Jaccard     & 25\% &\bestal{0.24}&\bestal{0.24} &\bestal{0.06}&\bestal{0.31} \\
&                & 50\% &\bestal{0.26}&\bestal{0.24} &\bestal{0.06}&\bestal{0.32} \\
&                & \multirow{2}{*}{{75\%}}& &        &\bestal{0.08}&\bestal{0.34} \\
& FD + FC        &      &\bestal{0.34}&\bestal{0.25} &             &              \\ \hline
	\multirow{12}{*}{{\rotatebox[origin=c]{90}{Recall@20}}}
& Jaccard        & 25\% &             &              &             &\baslin{0.13} \\
&                & 50\% &             &              &             &\baslin{0.18} \\
&                & 75\% &\baslin{0.12$^*$}&          &             &\baslin{0.20} \\
& EIR            & 25\% &\baslin{0.12}&\baslin{0.10} &\baslin{0.13}& \\ 
&                & 50\% &\baslin{0.11}&\baslin{0.10} &\baslin{0.11}& \\
&                & 75\% &             &\baslin{0.10} &\baslin{0.12}& \\ \cline{2-7}
& FD Jaccard     & 25\% &\bestal{0.18}&              &\bestal{0.23}& \\
&                & 50\% &\bestal{0.19}&              &\bestal{0.23}& \\
&                & 75\% &\bestal{0.14}&              &\bestal{0.20}& \\
&  FC + FD       & 25\% &             &\bestal{0.14} &             &\bestal{0.30} \\
&                & 50\% &             &\bestal{0.14} &             &\bestal{0.30} \\
&                & 75\% &             &\bestal{0.13} &             &\bestal{0.31} \\ \hline
	     \multirow{12}{*}{{\rotatebox[origin=c]{90}{DCG@20}}}
& ECP            & 25\% &\baslin{0.05}&             &             &        \\
&                & 50\% &\baslin{0.05}&             &             &        \\
&                & 75\% &\baslin{0.05}&             &             &        \\
& EIR            & 25\% &             &\baslin{0.06}&\baslin{0.05}& \baslin{0.12} \\
&                & 50\% &             &\baslin{0.06}&\baslin{0.05}& \baslin{0.12} \\
&                & 75\% &             &\baslin{0.06}&\baslin{0.05}& \baslin{0.12} \\  \cline{2-7}
& FD Jaccard     & 25\% &\bestal{0.10}&             &\bestal{0.11}&        \\
&                & 50\% &\bestal{0.11}&             &\bestal{0.11}&        \\
&                & 75\% &\bestal{0.08}&             &\bestal{0.10}&        \\
&  FC + FD       & 25\% &             &\bestal{0.08}&             &\bestal{0.17} \\ 
&                & 50\% &             &\bestal{0.08}&             &\bestal{0.17} \\ 
&                & 75\% &             &\bestal{0.08}&             &\bestal{0.17} \\ \hline
\end{tabular}
      \label{tab:exp_summary}
\end{table}

\begin{table} 
\caption[]{Experiments with combination of collaborative filtering for the least frequent (25\%) conditional items of the MovieLens data.}
\centering
    \begin{tabular}{lccc}
& MPR & Recall@20 & DCG@20 \\ \hline
Cosine			& 0.4978		& 0.0988 		& 0.0553	\\ 
Jaccard			& 0.4978		& 0.0988		& 0.0547	\\ 
ECP			& 0.4976 		& 0.0940 		& \baslin{0.0601}	\\ 
EIR			& \baslin{0.3203}	& \baslin{0.1291} 	& 0.0344	\\ 
FC Cosine		& 0.3583		& 0.1020		& 0.0505	\\ 
FD Cosine		& 0.2849		& 0.1578		& 0.0860	\\ 
FC Jaccard		& 0.3354		& 0.1770 		& \bestal{0.1031}	\\ 
FD Jaccard		& \bestal{0.2415}	& \bestal{0.1866} 	& 0.1010\\ 
FC ECP			& 0.2504		& 0.0940		& 0.0444\\ 
FD ECP			& 0.4212		& 0.1626 		& 0.0856\\ 
FC EIR			& 0.4125		& 0.0861 		& 0.0434\\ 
FD EIR			& 0.4529		& 0.1068		& 0.0560\\ \hline

%FC + FD Jaccard		& 0.3606 		& 0.1196		& 0.0709	\\\hline

%FC JC + Cosine		& 0.327		& 0.1515 	& 0.0785	\\ 
%FC JC + Jaccard	& 0.327		& 0.1515		& 0.0785	\\ 
%FC JC	+ ECP		& 0.391		& 0.0893		& 0.0778	\\ 
%FC JC	+ EIR		& 0.3902 	& 0.1084		& 0.0557	\\ \hline 
%FD JC + Cosine		& 0.3026 	& 0.0749	 	& 0.0423	\\ 
%FD JC + Jaccard	& 0.3026 	& 0.0749 	& 0.0423	\\ 
%FD JC	+ ECP		& 0.2385 	& 0.161		& 0.0416	\\ 
%FD JC	+ EIR		& 0.2805 	& 0.1307		& 0.0394\\ \hline
    \end{tabular}
      \label{tab:exp_infreq_comb}
\end{table}

\begin{table} 
\caption[]{Experiments on MovieLens with DBPedia content, all methods using Jaccard similarity.  FC and FD are the Fisher information methods of Sections~\ref{sect:FC} and~\ref{sect:FD} and multimodal uses the mix of both content and collaborative similarity.}
%JC and JS denote Jaccard and Jensen-Shannon, respectively, FC and FD are the Fisher information methods of Sections~\ref{sect:FC} and~\ref{sect:FD} and multimodal uses the mix of both content and collaborative similarity.}
\centering
    \begin{tabular}{lcc}
			& Recall@20 	& DCG@20 \\ \hline
Collaborative baseline	& 0.139		& 0.057 \\
Content baseline      	& 0.131 	& 0.056 \\
%JS collaborative 	& 	& \\
FC content 		& 0.239 	& 0.108 \\
%FC JS content 		& 		& \\ 
FD content		& 0.214 	& 0.093 \\
%FD JS content		& 		& \\
FC multimodal    	& 0.275 	& 0.123 \\ \hline 
%FD JS multimodal 	& 		& \\\hline
    \end{tabular}
      \label{tab:exp_content}
\end{table}

First, we give our results by relying on implicit item feedback information only, by using baseline similarity measures and Fisher models.  The results are summarized in Tables~\ref{tab:exp_summary}--\ref{tab:exp_infreq_comb} and explained in the sequel.
First we describe the parameter settings of the Fisher information methods.
The sample set of items as defined in Figs.~\ref{fig:pairwise}--~\ref{fig:multi} is the starting point of all models.
We have to be careful with selecting the sample items to avoid problems caused by sparse items.  For this purpose, we choose the most popular items in the training set as elements for the sample set.
Next  we investigate the effect of the sample size.
As we can see in Fig.~\ref{fig:samples}, recommendation quality saturates at a certain sample set size. Therefore we set the size of the sample set to $20$ for the remaining experiments. 

As expected, the choice of the distance function strongly affects the performance of the Fisher models. As seen in Table~\ref{tab:exp_infreq_comb}, the overall best performing distance measure is Jaccard for both types of Fisher models. 
The results in Table~\ref{tab:exp_summary} show that the linear combination of the standard normalized scores of the Fisher methods outperforms the best unimodal methods (Fisher with Jaccard) in 
case of the Netflix and the Books data sets, while for the MovieLens and Yahoo!\ Music datasets the Fisher distance with Jaccard performs best. 

One of the main challenges in the field of recommendation systems is the ``cold start'' problem. Since one of the main reasons behind the item-to-item 
recommendation is the assumption of unknown or ``cold'' users, we examine the performance in case of infrequent items. Figure~\ref{fig:supp_netf} shows the superiority of the Fisher methods for
low item support.  As support increases, best results are reached by blending based on item support.  If the current session ends with an item of high support, we may take a robust baseline recommender.  And if the support is lower, less than around 100, Fisher models can be used to compile the recommendation.

In Fig.~\ref{fig:sparsity} we can see the item-to-item similarity values between the least frequent and between the most frequent items for different models. The feedback-based similarity may result arbitrary 
ranking for the least frequent items in comparison to the Fisher based similarity using the same modality. As expected the density of the content-based similarity models are more balanced. 

Finally we turn to our content based and multimodal, mixed content and feedback recommendation experiments.
In Table~\ref{tab:exp_content} we show our experiments with DBPedia content as a modality on MovieLens. Due the complexity of the corresponding RDF graphs, we set the size of the sample set for both Fisher models to 10. 
The overall best performing model is the multimodal Fisher with Jaccard similarity, while every unimodal Fisher method outperform the baselines. 

Our experiments clearly show that the Fisher information based methods can blend different modalities such as content and feedback without the need of setting external parameters or applying learning for blending.  Note that for the results of Table~\ref{tab:exp_content}, we used the method of equation \eqref{eq:fk_multi} with no additional steps required for fusing the modalities.

\section{Conclusions}

In this paper, we considered the session based item-to-item recommendation task, in which the recommender system has no personalized knowledge of the user beyond the last items visited in the current user session.
We proposed Fisher information based global item-item similarity models for this task.
We reached significant improvement over existing methods by experimenting with a variety of data sets as well as evaluation metrics. In particular, by making the best effort to 
reproduce the experiments of \cite{koenigstein2013towards}, we showed that our method is superior in all aspects, even if we only use the implicit item ratings for recommendation.

We consider our results for simple, non-personalized item-to-item recommendation as first step towards demonstrating the power of the method.
As a key feature, we are able to fuse different modalities including collaborative filtering, content and side information, without the need for learning weight parameters or using wrapper methods.
In the future, we plan to extend our methods to personalized recommendation settings.

We made our experiments fully reproducible by releasing the source code of not just the algorithm, but also the preprocessing steps taken over the standard 
experimentation data sets used in our measurements.

\section{Acknowledgments}
The publication was supported by the PIAC\_13-1-2013-0205 project of the Research and Technology Innovation Fund, by the Momentum Grant of the Hungarian Academy of Sciences and by the Mexican Postgraduate Scholarship of the Mexican National Council for Science and Technology (CONACYT). 
 
\bibliographystyle{abbrv}

\bibliography{i2i,simker_ref,latex-common/bib/spam,latex-common/bib/recsys.bib,latex-common/bib/telco-blog,latex-common/bib/www,latex-common/bib/sketch-ppr,latex-common/bib/image,latex-common/bib/bicluster,latex-common/bib/clef2009,latex-common/bib/svd-missing.bib,session-drop,references}

\end{document}